\documentclass[seceq]{ptptex}

\usepackage{wrapft}
\usepackage{graphicx}

\notypesetlogo  %comment in if to eliminate PTPTeX logo
\title{%        %You can use \\ for explicit line-break.
Anisotropic flow and other collective phenomena measured in Pb-Pb collisions with ALICE at the LHC
}
\author{%       %Use \scshape  for the family name.
Ilya \textsc{Selyuzhenkov} for the ALICE Collaboration
}

\inst{%         %Affiliation, neglected when [addenda] or [errata].
Research Division and ExtreMe Matter Institute EMMI,\\
GSI Helmholtzzentrum f\"ur Schwerionenforschung, Darmstadt, Germany
}

\abst{%         %This abstract is neglected when [addenda] or [errata].
Recent results of the anisotropic flow measurements
by the ALICE Collaboration at the LHC are reviewed.
Directed, elliptic, triangular, and quadrangular flow
are presented differentially vs. transverse momentum,
pseudo-rapidity, and the collision centrality
for charged and identified particles.
Experimental probes of local parity violation using the charge dependent azimuthal
correlations with respect to the reaction plane are also discussed.
}

\begin{document}
\maketitle

\section{Introduction}
An azimuthal anisotropic flow describes a collectivity among particles
produced in heavy-ion collision, and it is recognized as
one of the key observable which provides information on
the early time evolution of the nuclei interaction.
This ISMD2011 Conference proceedings highlight recent results
by the ALICE Collaboration from the anisotropic flow measurements
for Pb-Pb collisions at \mbox{$\sqrt{s_{\rm NN}}$ = 2.76 GeV}.
Current status of probes of parity symmetry violation in strong interaction
using the charge dependent azimuthal correlations
with respect to the reaction plane is also discussed.

\section{Anisotropic flow, fluctuations, and non-flow correlations}
Anisotropic transverse flow is usually quantified by the coefficients
(harmonics) in the Fourier decomposition of the azimuthal distribution
of particles with respect to the reaction plane.
The collision reaction plane, which is defined by
the impact parameter and the colliding nuclei direction,
is not known experimentally
and the anisotropic flow coefficients can be only extracted
from azimuthal correlations between produced particles
(for review of the anisotropic flow measurement techniques see \cite{Voloshin:2008dg}).
The main challenge in the anisotropic flow measurement
is to disentangle contribution from correlations
not related to the reaction plane (so called non-flow correlations),
and to understand the impact on the measured flow
from the event-by-event fluctuations
(e.g. due to fluctuating energy density
in the overlap zone of two nuclei).

\begin{figure}[ht]
\begin{center}
  \includegraphics[width=.5\textwidth]{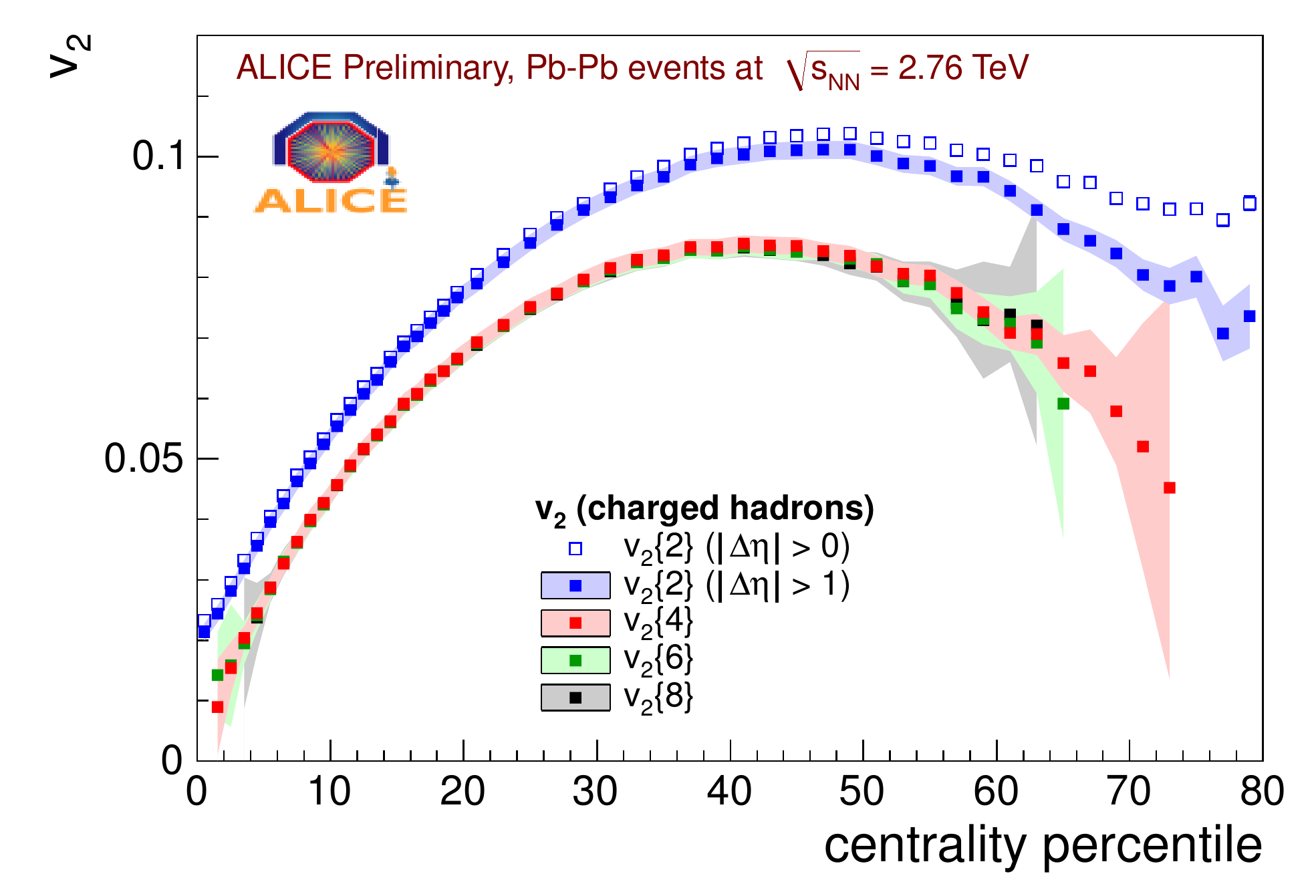}%
  \includegraphics[width=.5\textwidth]{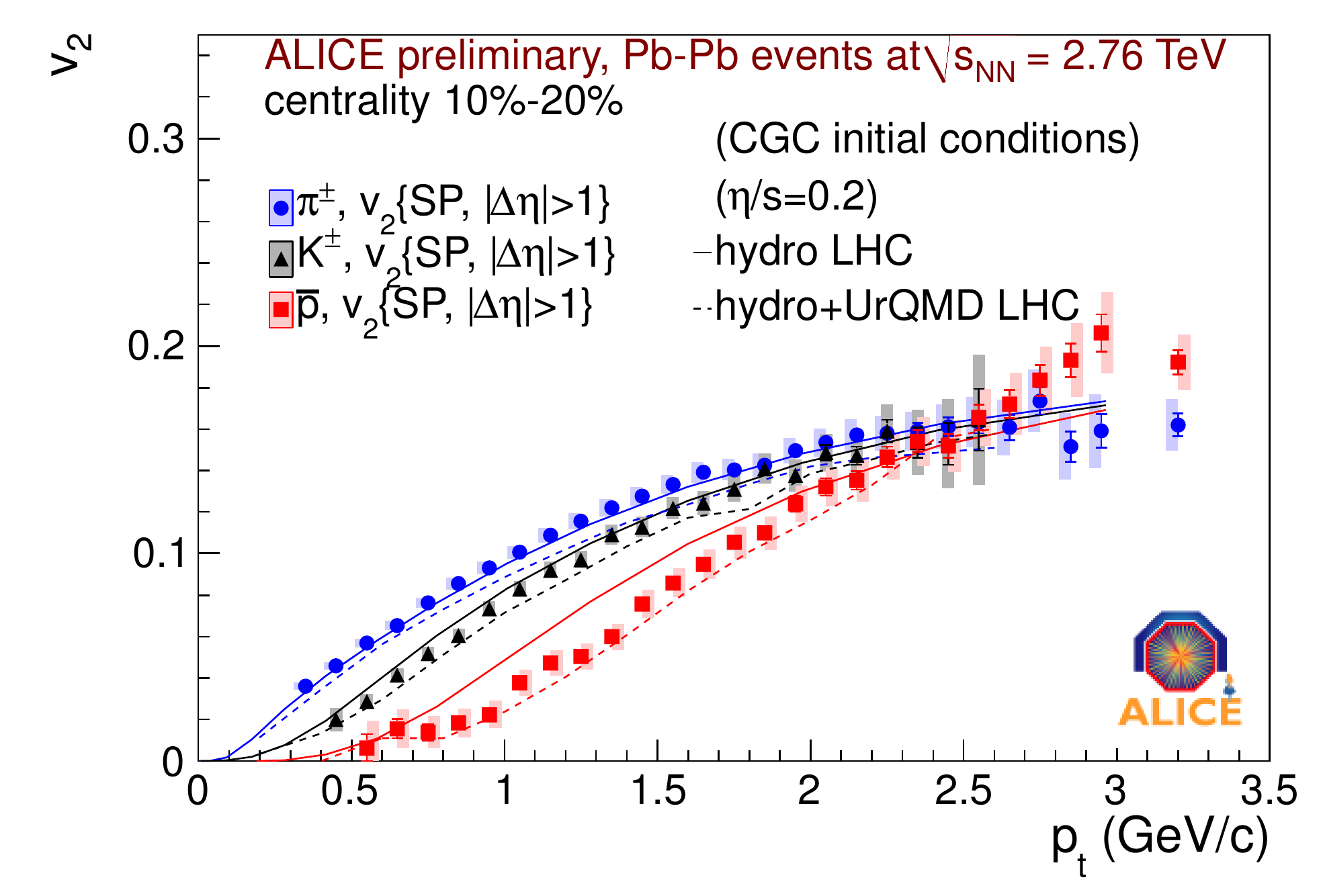}%
{\mbox{}\\\vspace{-0.2cm}
\hspace{1cm}\mbox{~} \bf (a)
\hspace{+6.7cm}\mbox{~} \bf (b)}
{\mbox{}\\\vspace{-0.1cm}}
\caption{Elliptic flow, $v_2$, measured for Pb-Pb collisions at $\sqrt{s_{\rm NN}} = $~2.76~TeV.
(a) $v_2$ of charged particles vs.~centrality, (b) $v_2$ vs.~transverse momentum for pions, kaons, and anti-protons.
Figure~(a) taken from \cite{Bilandzic:2011ww} and figure (b) {from \cite{Krzewicki:2011ee}}.
}
  \label{fig:1}
\end{center}
\end{figure}

Figure \ref{fig:1}(a) shows a systematic study of non-flow and flow fluctuations
for the elliptic flow, $v_2$, measured
for Pb-Pb collisions at  \mbox{$\sqrt{s_{\rm NN}}$ = 2.76 GeV}.
The magnitude of non-flow effects is driven by the difference between $v_2$
estimated from the two-particle azimuthal correlations without ($v_2\{2,|\Delta \eta|>0\}$)
and with ($v_2\{2,|\Delta \eta|>1\}$) pseudo-rapidity separation between correlated particles
which greatly suppress non-flow effects from short-range correlations.
Flow fluctuations can be estimated from the difference between
the results from two-particle correlations with a large
rapidity gap and those from multi- (4, 6, and 8) particle cumulants
(for an estimate of the elliptic flow fluctuations under the assumption of the
small or Gaussian fluctuations see \cite{Collaboration:2011yb}).
The results in Fig.~\ref{fig:1}(a) show that both flow fluctuations
and non-flow (mainly for the peripheral collisions)
are significant and have to be seriously taken into account when comparing
measured anisotropic flow with theoretical calculations.

\section{Elliptic flow of identified particles}
Since the success of the ideal hydrodynamic description of the elliptic flow, $v_2$, for
the central Au-Au collisions at {RHIC \cite{Ackermann:2000tr}},
the hydrodynamics is considered as the
most appropriate theory to describe a thermalized phase
in the time evolution of the system created in a heavy-ion collision.
An important test of the hydrodynamic description at the LHC is the interplay
between radial (azimuthally symmetric radial expansion)
and anisotropic flow which result in the mass splitting
of the elliptic flow at small transverse momenta.
Figure~\ref{fig:1}(b) shows the elliptic flow of pions, kaons, and anti-protons
vs. particle transverse momenta, $p_{\rm t}$, measured with the scalar product\cite{Voloshin:2008dg} (SP) method.
The mass dependence of $v_2$ at low transverse momenta, $p_{\rm t} < 2.5$~GeV/c, is
well reproduced by viscous hydrodynamic model calculations \cite{Shen:2011eg}
with a color glass condensate initial condition
(solid lines in Fig.~\ref{fig:1}(b)).
Agreement with data, especially for protons,
is improved when adding a hadronic cascade phase
into the model calculations (dashed lines in Fig.~\ref{fig:1}(b)).
\begin{figure}
\begin{center}
  \includegraphics[width=.5\textwidth]{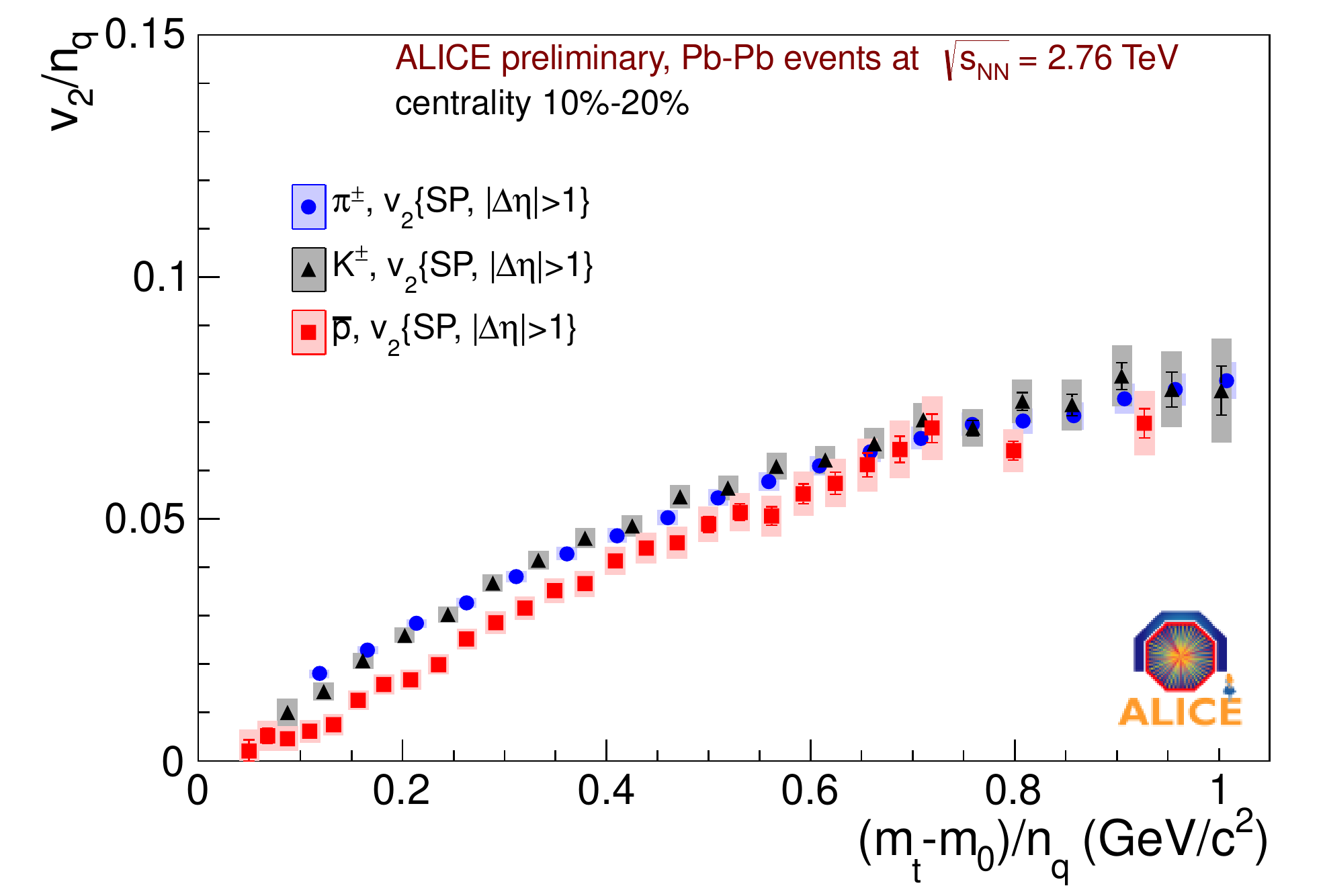}%
  \includegraphics[width=.5\textwidth]{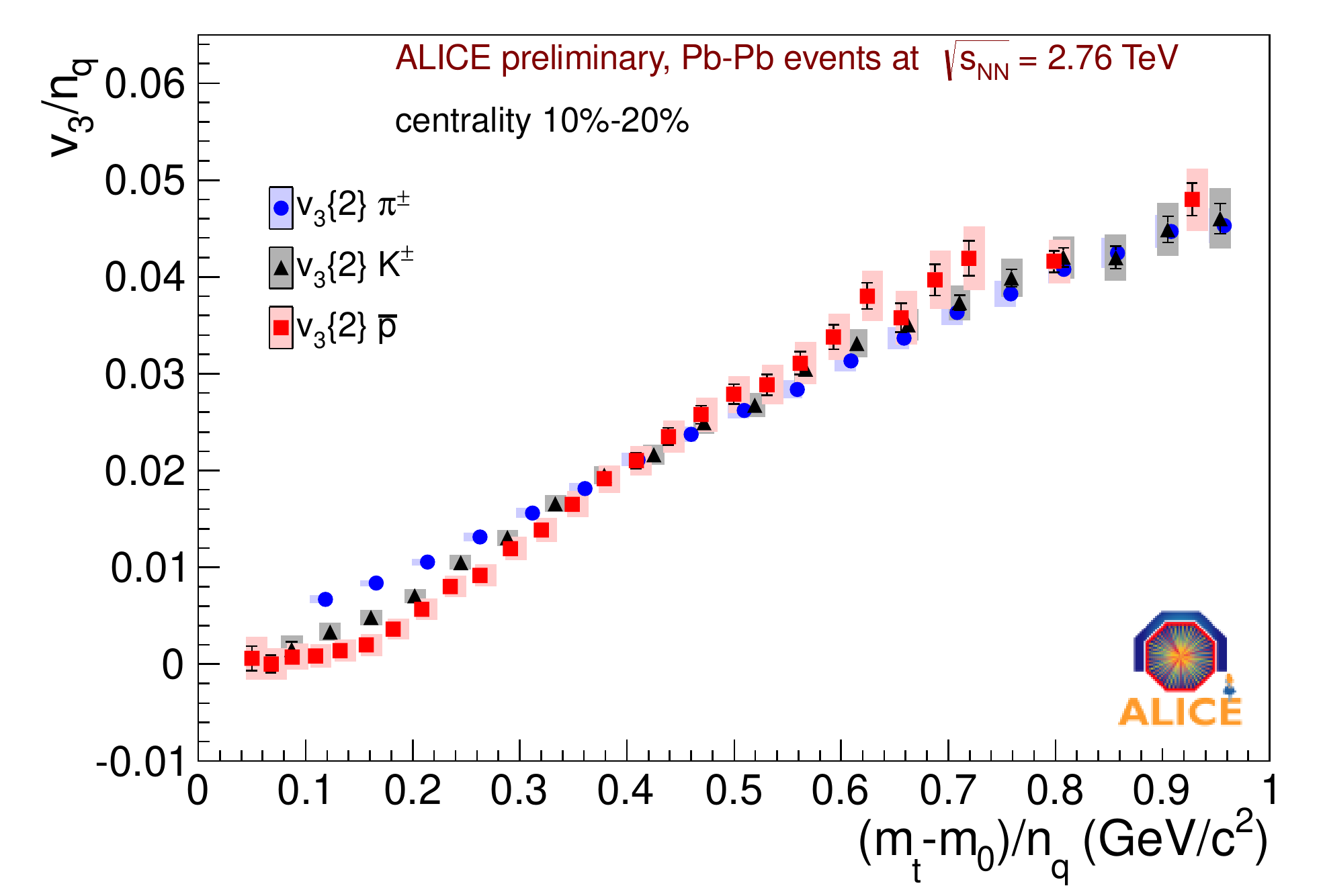}%
{\mbox{}\\\vspace{-0.2cm}
\hspace{1cm}\mbox{~} \bf (a)
\hspace{+6.7cm}\mbox{~} \bf (b)}
{\mbox{}\\\vspace{-0.1cm}}
  \caption{(a) Elliptic, $v_2$,  and (b) triangular, $v_3$, flow measured with the scalar product (SP) method
for Pb-Pb collisions at $\sqrt{s_{\rm NN}} = $~2.76~TeV.
Elliptic and triangular flow are scaled with the constituent number of quarks and plotted vs. transverse kinetic energy per quark.
Figures taken {from \cite{Krzewicki:2011ee}}.}
  \label{fig:2}
\end{center}
\end{figure}
Figure~\ref{fig:2}(a) shows elliptic flow of pions, kaons, and anti-protons
scaled with the number of constituent quarks, $n_q$ ($n_q=2$ for mesons, and $n_q=3$ for baryons),
vs. transverse kinetic energy per quark, $(m_t - m_0)/n_q$.
The observed approximate scaling of $v_2$ with the number of quarks
in the range of \mbox{$p_{\rm t} \sim 2-3$ GeV/c} (\mbox{$m_{\rm t} \sim 0.6-1.0~{\rm GeV/c^2}$})
reflects collectivity at the quark level and suggest
that the system evolved through the phase of deconfined quarks and gluons.

\section{Triangular and higher harmonic flow}
Recent progress in understanding  the connection
between the anisotropic flow and the fluctuations of the
energy density in the initial state of the heavy-ion collision
showed that not only the dominant elliptic flow component is important,
but that other harmonics such as triangular flow are crucial
for the realistic description of the system created during the collision (see \cite{:2011vk} and references therein).
\begin{figure}
\begin{center}
  \includegraphics[width=.5\textwidth]{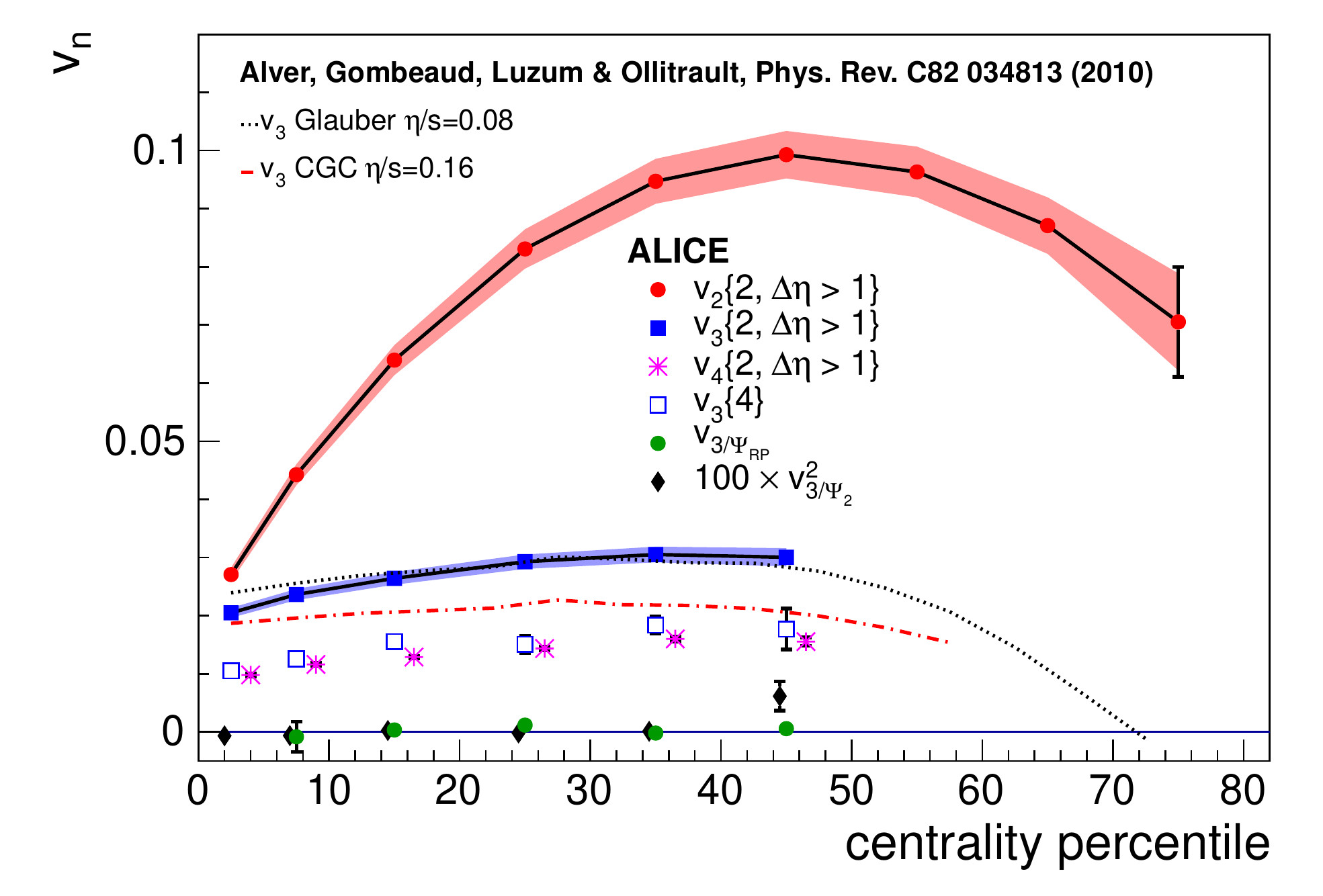}%
  \includegraphics[width=.45\textwidth]{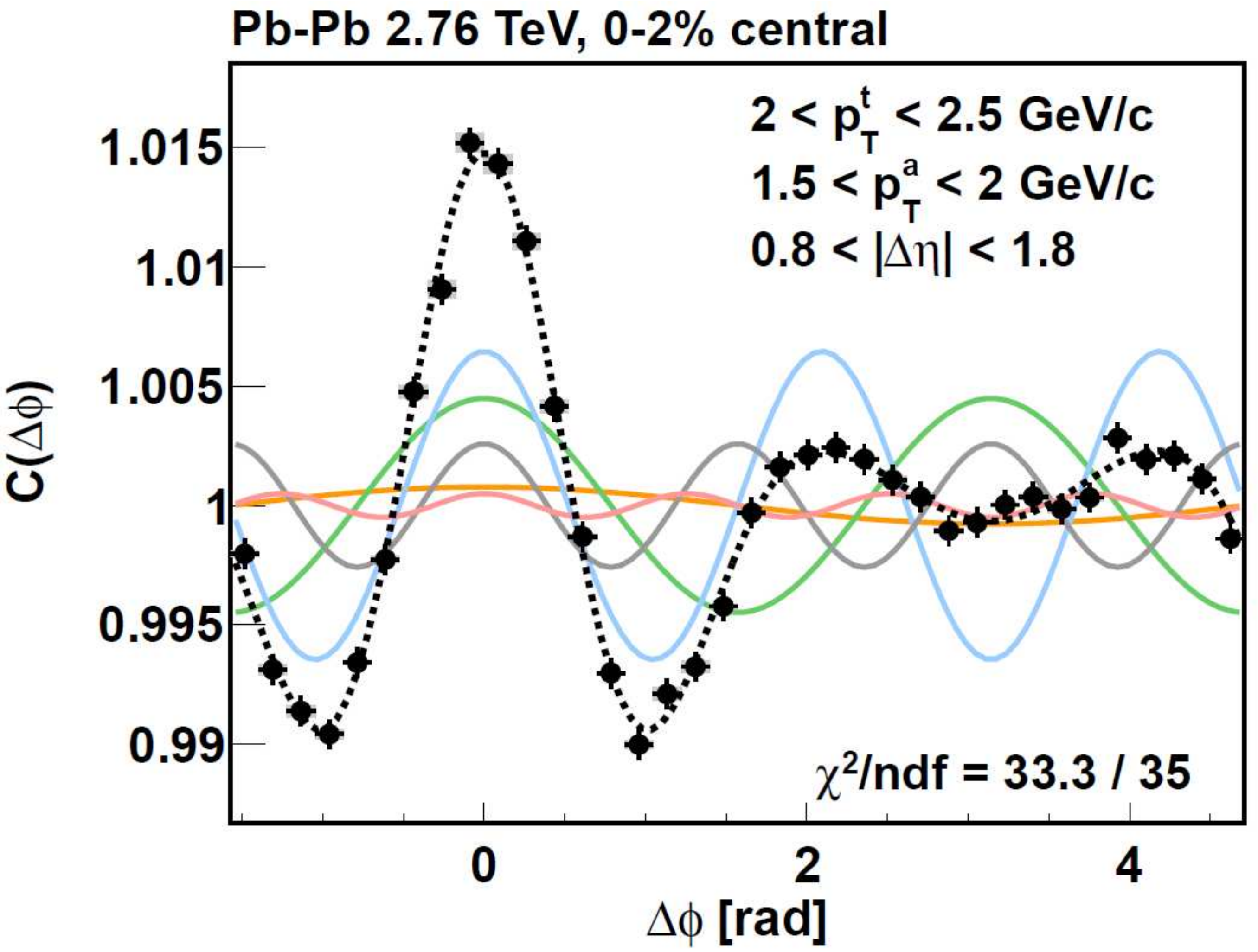}%
{\mbox{}\\\vspace{-0.cm}
\hspace{1cm}\mbox{~} \bf (a)
\hspace{+6.7cm}\mbox{~} \bf (b)}
{\mbox{}\\\vspace{-0.1cm}}
  \caption{(a) 
Elliptic, triangular, and quadrangular flow 
vs. collision centrality measured for
Pb-Pb collisions at $\sqrt{s_{\rm NN}} = $~2.76~TeV. (b) Two-particle azimuthal correlations 
measured with large ($|\eta|>1$) rapidity separation between particles and their decomposition into
anisotropic flow harmonics.
Figure~(a) taken from \cite{:2011vk,Collaboration:2011yb} and figure (b) {from \cite{Aamodt:2011by}}.}
  \label{fig:3}
\end{center}
\end{figure}
Figure~\ref{fig:3}(a) shows elliptic, triangular, and quadrangular flow 
vs. collision centrality measured with two- and four-particle correlations for
Pb-Pb collisions at $\sqrt{s_{\rm NN}} = $~2.76~TeV.
Measured triangular flow, $v_3$, behaves as it is expected for collective correlations
from the fluctuations of the initial geometry, i.e. i) weak centrality dependence
follows calculations with fluctuating initial condition 
(solid blue squares vs. dotted black line),
ii) ``proper" ratio of $v_3\{2\}/v_3\{4\}\approx 2$ which is expected for the case of pure {fluctuations \cite{Bhalerao:2011yg}},
and iii) no correlation of $v_3$  with respect to the true reaction plane
($v_3$ measured with the reaction plane estimated from deflection of
neutron spectators is consistent with zero, see green points in Fig.~\ref{fig:3}(a)),
iv) no correlation between $v_3$ and the azimuthal modulations
in the second flow harmonic, $v_2$
(black diamonds in Fig.~\ref{fig:3}(a)).
Another evidence for the collective origin of the triangular flow
is the similar mass splitting and the number of quark scaling to that
of elliptic flow which is demonstrated in Fig~\ref{fig:2}(b).
Note that in contrast to the $v_2$ results in Fig.~\ref{fig:2}(a), there
is no pseudo-rapidity separation between correlated particles
in $v_3$ measurements which may results in the additional bias
at small $(m_t - m_0)/n_q$ values.

It is remarkable that including higher-order flow harmonics allows to reproduce
the ``ridge" and ``Mach-cone" features of the two-particle azimuthal correlations
at low transverse momenta (see Fig.~\ref{fig:3}(b)),
which were originally interpreted as results of the
propagation of the hard probe (e.g. jet) through the dense medium.

\section{Directed flow}
Directed flow, $v_1$, is sensitive to the earliest, pre-equilibrium,
times in the evolution of the system (see \cite{Voloshin:2008dg} and references therein).
\begin{figure}
\begin{center}
  \includegraphics[width=.5\textwidth]{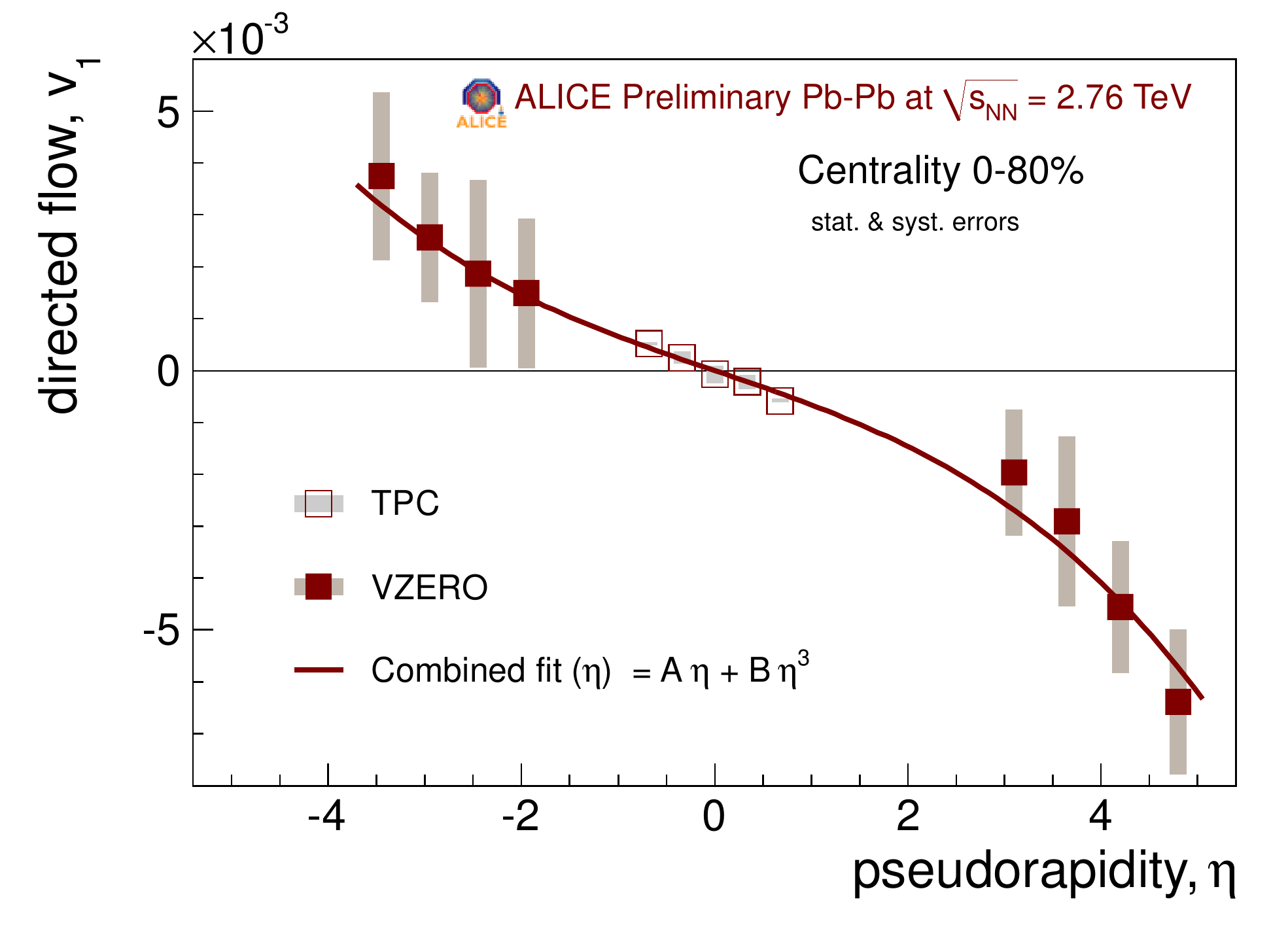}%
  \includegraphics[width=.5\textwidth]{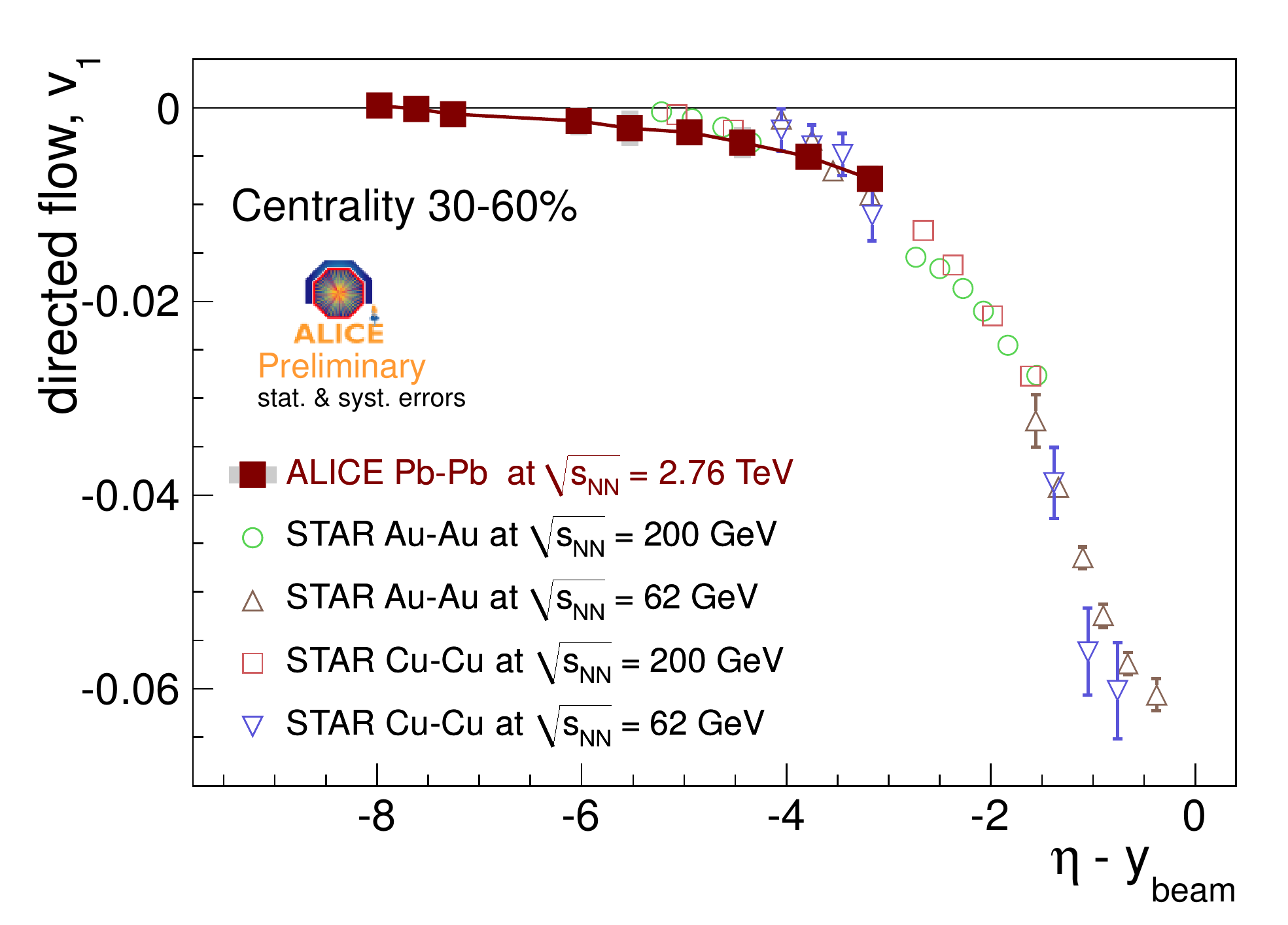}%
{\mbox{}\\\vspace{-0.2cm}
\hspace{1cm}\mbox{~} \bf (a)
\hspace{+6.7cm}\mbox{~} \bf (b)}
{\mbox{}\\\vspace{-0.1cm}}
  \caption{Directed flow, $v_1$, for Pb-Pb collisions at $\sqrt{s_{\rm NN}} = $~2.76~TeV.
(a) $v_1$ over large rapidity range, $|\eta|<5.1$. (b) longitudinal scaling of $v_1$.
Figures taken {from \cite{Selyuzhenkov:2011zj}}.}
  \label{fig:4}
\end{center}
\end{figure}
Figure \ref{fig:4}(a) shows $v_1$ of charged particles
measured in a wide rapidity range with the reaction plane estimated from the deflection of the
spectator neutrons at beam rapidity.
The absolute sign of directed flow is fixed in the measurement by the same
convention as used at RHIC, i.e.~spectators with $\eta>0$ possess a positive $v_1$.
The measured negative slope of $v_1$ as a function of pseudo-rapidity is
opposite to the predictions for LHC energies from
the quark-gluon string model with
parton rearrangement \cite{Bleibel:2007se}
and fluid dynamical calculations \cite{Csernai:2011gg}
which suggest a much stronger signal with
a positive slope of $v_1$.
Figure~\ref{fig:4}(b) shows $v_1$ measured as a function of beam rapidity
which is consistent with the longitudinal scaling previously observed at RHIC energies.

\section{Probes of local parity violation in strong interaction}
The extreme magnetic field created during a non-central relativistic heavy-ion collision
may spontaneously excite instantons and sphalerons from the QCD vacuum
which violates parity symmetry of the strong interactions.
It is predicted by Kharzeev {\it et al.} \cite{Kharzeev:2007jp} that this may
result in the experimentally observable separation of charges along the magnetic field.
Since the magnetic field is perpendicular to the collision reaction plane,
Voloshin \cite{Voloshin:2004vk} proposed to use the anisotropic flow measurement
technique to experimentally probe the effects of charge separation.
\begin{figure}
\begin{center}
  \includegraphics[width=.5\textwidth]{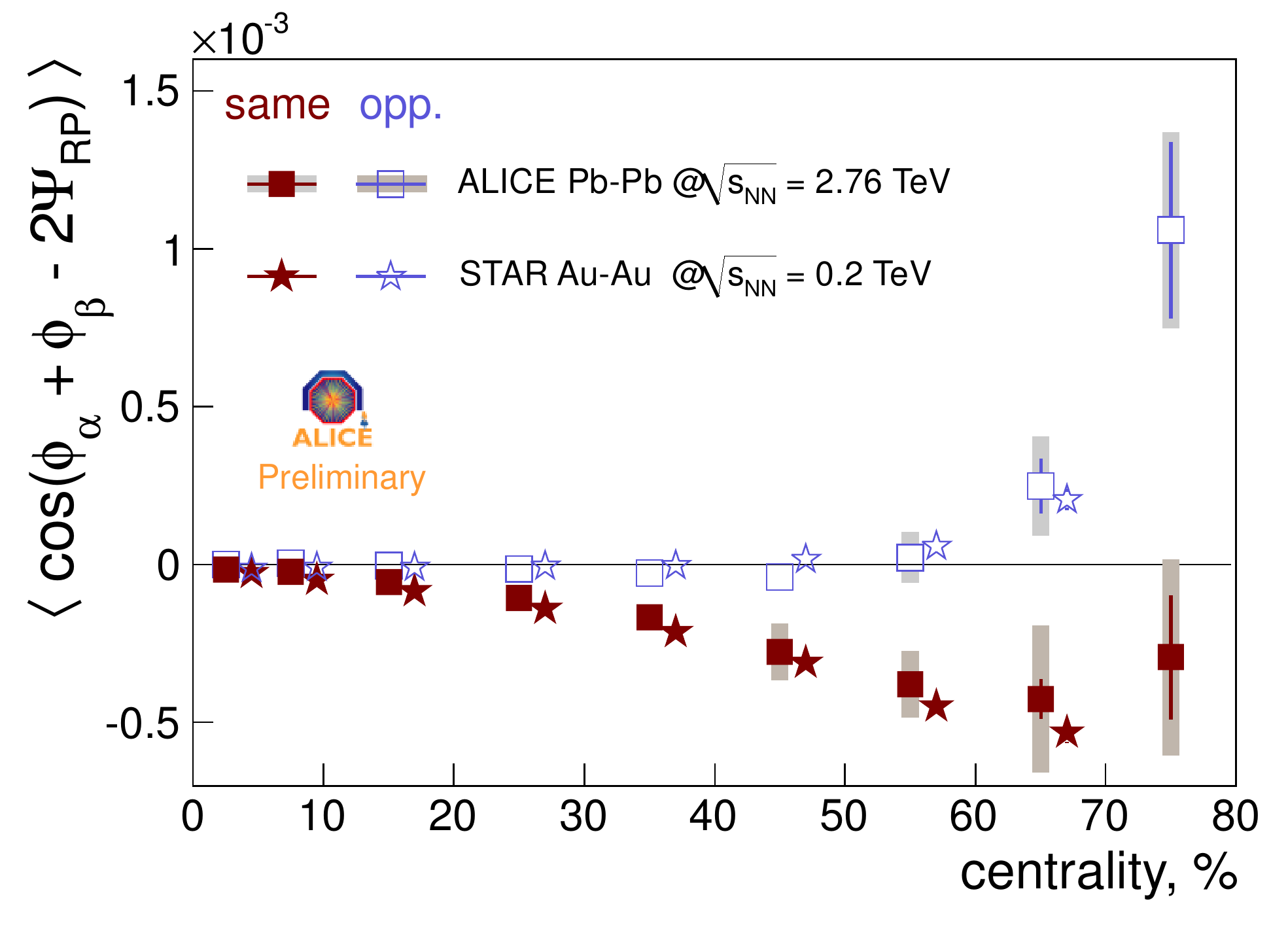}%
  \includegraphics[width=.5\textwidth]{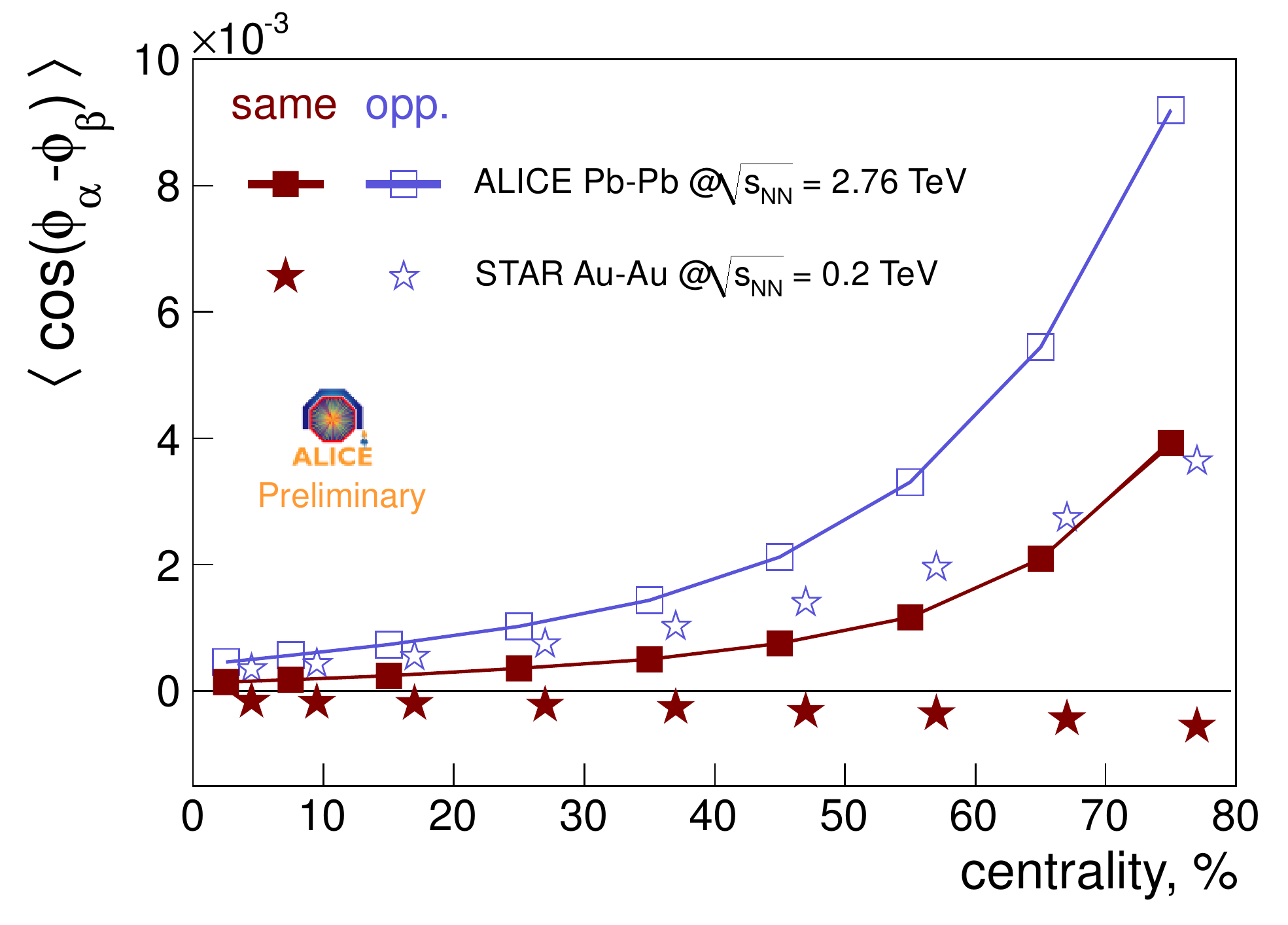}%
{\mbox{}\\\vspace{-0.2cm}
\hspace{1cm}\mbox{~} \bf (a)
\hspace{+6.7cm}\mbox{~} \bf (b)}
{\mbox{}\\\vspace{-0.1cm}}
  \caption{
Charged dependent azimuthal correlations  vs. centrality measured for Pb-Pb collisions at $\sqrt{s_{\rm NN}} = $~2.76~TeV
and Au-Au collisions at $\sqrt{s_{\rm NN}} = $~0.2~TeV:
(a)~two-particle correlations with respect to the reaction plane,
(b)~1st harmonic two-particle correlations.
Figures adapted {from \cite{Collaboration:2011sm}}.
}
  \label{fig:5}
\end{center}
\end{figure}
Figure~\ref{fig:5}(a) shows the experimental results
for the charge-dependent two-particle correlation with respect to the reaction plane:
$\langle \cos(\phi_{\alpha}+\phi_\beta- 2\Psi_{\rm RP})\rangle$, where $\phi_{\alpha,\beta}$ is the azimuthal angle and $\alpha,\beta$ charge of the particle, and $\Psi_{\rm RP}$ is the reaction plane angle.
Clear charge separation is observed at both RHIC and LHC energies
with a very similar magnitude and centrality dependence of the correlations.
The $\langle \cos(\phi_{\alpha}+\phi_\beta- 2\Psi_{\rm RP})\rangle$
observable has direct sensitivity to the event-by-event charge fluctuations,
but it is parity even and thus is sensitive to effect unrelated to the symmetry violation.
The presence of parity even background correlations
which contributes to the measured charge separation
at RHIC and LHC significantly complicates the interpretation of the data.
Among possibly large contributions from the parity even backgrounds
are flow fluctuations in the first flow harmonic \cite{Teaney:2010vd}
and effects of local charge {conservation \cite{Pratt:2010gy}}.
Figure~\ref{fig:5}(b) presents the 1st harmonic two-particle correlations,
$\langle \cos(\phi_{\alpha}-\phi_\beta)\rangle$,
which in contrast to the $\langle \cos(\phi_{\alpha}+\phi_\beta- 2\Psi_{\rm RP})\rangle$
show opposite sign at LHC than at RHIC for the same charge correlations.
Results for the $\langle \cos(\phi_{\alpha}-\phi_\beta)\rangle$ correlator
are dominated by the parity conserving background sources
and this may provide additional insights on the origin of the measured charge separation.
Currently, large theoretical uncertainties
in the estimate of background correlations as well as lack of
quantitative predictions from the models which incorporate parity
symmetry violation in QCD make 
the data interpretation difficult and further theoretical
developments in this direction are extremely important.

\section{Summary and outlook}
The anisotropic flow harmonics up to the fifth order
have been measured for Pb-Pb collisions
at $\sqrt{s_{\rm NN}}$ = 2.76 GeV by the ALICE Collaboration at the LHC.
Altogether, directed, elliptic, triangular, and quadrangular flow measurements
provide strong constraints on the properties of the system created
during the heavy-ion collision such as
viscosity, initial conditions, and the equation of state.
Charge separation of particles with respect to the collision reaction plane,
which was first observed at RHIC energies, is
now measured for Pb-Pb collisions
at $\sqrt{s_{\rm NN}}$ = 2.76 GeV by the ALICE Collaboration at the LHC.
The charge-dependent two-particle azimuthal correlations
with respect to the reaction plane are very similar to that at RHIC energies,
while the background dominated 1st harmonic
two-particle azimuthal correlations 
show a different sign at LHC than at RHIC for the same charge correlations.
This provides strong experimental constraints
on the possible mechanism of the measured charge separation.

\section*{Acknowledgements}
This work was supported by the Helmholtz Alliance
Program of the Helmholtz Association, contract HA216/EMMI ``Extremes of
Density and Temperature: Cosmic Matter in the Laboratory".

\end{document}